\documentclass[aps,prc,twocolumn,floatfix,showpacs,a4paper,
nofootinbib,amsmath,amssymb]{revtex4-2}
\usepackage{graphicx}
\usepackage{dcolumn}
\usepackage{bm}
\usepackage{color}
\usepackage[utf8]{inputenc}

\newcommand{\be}{\begin{equation}}
\newcommand{\ee}{\end{equation}}
\newcommand{\ba}{\begin{eqnarray}}
\newcommand{\ea}{\end{eqnarray}}
\newcommand{\bd}{\begin{displaymath}}
\newcommand{\ed}{\end{displaymath}}
\renewcommand{\vec}[1]{\mbox{\boldmath$#1$}}

\begin{document}
\title{Radiation pattern and source size of particles in nanoplasmonic fusion}

\author{ 
L.P. Csernai$^{1,2,3,4}$, 
T. Cs\"org\H o$^1$, I. Papp$^{1,5,6}$, M. Csete$^{1,7}$,  \\ 
J.P. Hansen$^2$, A. Szenes$^{1,7}$, K. Tamosiunas$^8$, 
D. Vass$^{1,7}$, T.S. Bir\'o$^1$, \\
M. Veres$^1$ and N. Kro\'o$^1$ (part of NAPLIFE and FUSENOW Collaborations)}

\affiliation{\medskip
$^1$ HUN-REN Wigner Research Centre for Physics, Budapest, Hungary\\
\mbox{$^2$ Department of Physics and Technology, University of Bergen, Norway}\\
\mbox{$^3$ Frankfurt Institute for Advanced Studies, Frankfurt/M, Germany}\\
\mbox{$^4$ Csernai Consult Bergen, Bergen, Norway}\\
\mbox{$^5$ HUN-REN Centre for Energy Research, Budapest, Hungary}\\
\mbox{$^6$ King's College the University, London, United Kingdom}\\
\mbox{$^7$ Department of Optics and Quantum Electronics, University
of Szeged, Szeged, Hungary}\\
\mbox{$^8$ Vytautas Magnus University, Vileikos 8, Kaunas, Lithuania}\\
}
\vskip -4mm

\begin{abstract}
For the angular radiation patterns of proton, deuteron or alpha emission 
we present a way using particle-in-cell simulation of laser induced 
nanoplasmonic fusion.
The differential Hanbury-Brown and Twiss analysis is widely used 
in astrophysics and in relativistic heavy ion physics to determine 
the source size of emitted particles.  Here, we show how this method 
could be adopted for inertial confinement fusion.
This method aims to determine the parameters of emitted nuclei 
after the fusion target ignition. 
In addition to spatial volume, the method can detect specific 
space-time correlation patterns connected to the collective 
flow post-ignition.
In the NAPLIFE project our aim is to avoid thermalization and fluidization 
as much as possible at each stage of the fusion process. 
As the original laser beam is non-thermal and not equilibrated in any way 
it is obvious that we can minimize energy loss if we exploit the initial 
available energy in a non-thermal way. The detailed dynamics of deuterium and 
alpha production is not aimed at and not addressed by this paper. 
\end{abstract}



\maketitle

\section{Introduction}
\label{Intro}

The two-particle correlation function analysis was presented recently
in a brief {\it Communication} for the purpose of studying
the size and dynamics of ignition in NAnoPlasmonic Laser Induced 
Fusion Energy (NAPLIFE) project
\cite{Cs-Univ-2024}.
The Hanbury Brown and Twiss (HBT) analysis, stemming from astrophysics
is a combination of standard two particle correlation functions; it
is adequate to analyze the size, timespan, of hadronization in
relativistic heavy ion collisions
\cite{WF10}
and even collective flow
\cite{CsV2014, CsVW2014}
It is based on the Boson's wave function properties.

Nanotechnology was considered in recent years to facilitate
laser induced fusion. These attempts aimed for reducing laser
beam reflection and increasing absorption by long (5-10 $\mu$m)
nanowires, which were parallel to the direction of laser 
irradiation
\cite{Kay1,Kay2,Roc1,Bon1,Bon2}.

In contrast the NAPLIFE project uses resonant nanoantennas,
as half wavelength dipoles, which are orthogonal to the
laser irradiation.
Laser induced nanoplasmonic fusion reactions even at very
low laser pulse energy of 30 mJ showed the creation of
deuterium atoms and even He$^4$ atoms in small quantities.
The nuclei of these particles are bosons
\cite{Kroo23,Rigo19,Kumari22, Racz22, Cs-PRE-23}. 
These bosons with the same quantum structure are expected to show the 
same correlation properties as the emitted particles at
high energies.

\begin{figure}[!h]
\includegraphics[width=75mm]{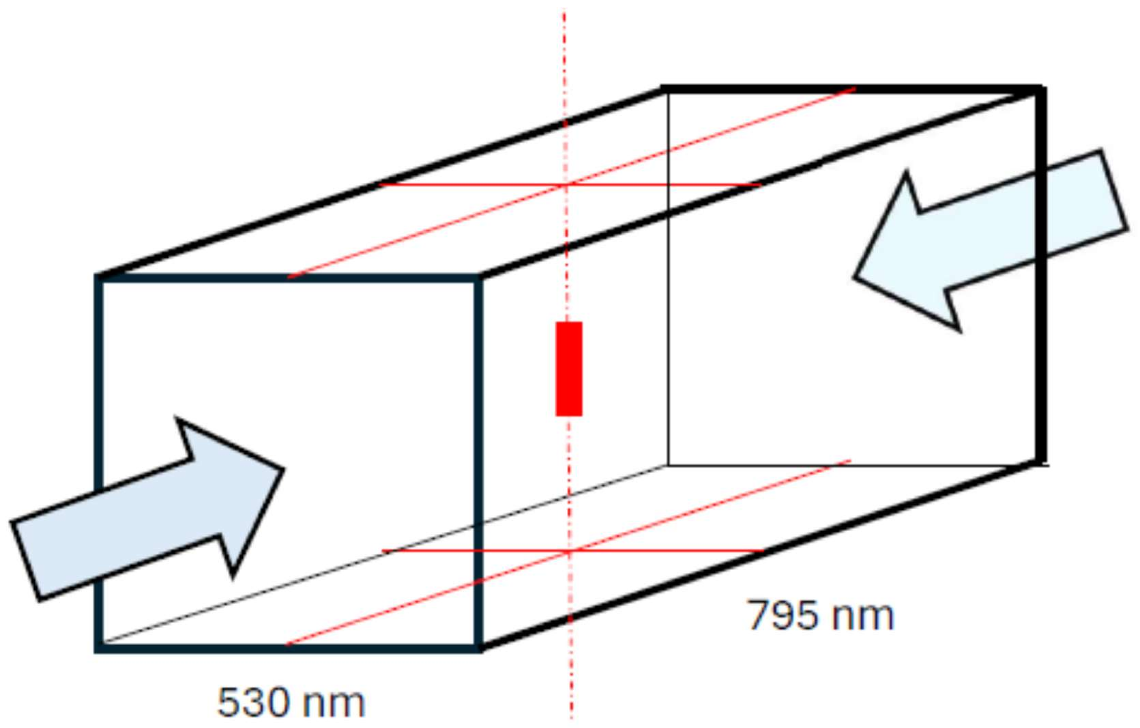}
\vskip -2mm
\caption{
(color online) Schematic view of the calculation box (CB) and the transverse 
orientation of nanorod antenna (red, x-direction) relative to the two-sided 
laser irradiation (grey arrows, z-direction). The CB is divided into cubic 
cells in the PIC method. Lagrangian fluid cells of different particle 
species are represented by different types of marker particles. The 
length/thickness 
of nanorod antenna is resonant to the laser light frequency in the material of 
fusion fuel target. Conduction electrons are resonating within the 
nanorod antenna 
in orthogonal direction to the laser beam direction. The protons 
neighboring the 
nanorod antenna are attracted by the resonating electrons and moving parallel 
to the nanorod. I.e. in this mechanism the accelerated protons are 
orthogonal to 
the laser beam, in contrast to the usual TNSA configuration.
}
\vskip -1mm
\label{a0}
\end{figure}

Proton and heavier ion acceleration by intensive laser beam is
well studied by the ponderomotive force and by the Target Normal
Sheath Acceleration (TNSA). Using these techniques even medical cancer
treatment was envisaged for quite some time.  Recently the unique
method was proposed to apply resonant plasmonic nano-rod antennas for the
regulation of local absorption of the laser beam energy and momentum.  
This method by the NAPLIFE collaboration
\cite{CsEA2020}
opened new possibilities.
With two-sided irradiation 
{\color{blue}
(Fig. \ref{a0})
} 
in one dimensional
geometry it will ensure rapid, stable ignition avoiding the
Rayleigh-Taylor instability
\cite{CsKP18}.
%
%
%
Using the Laser Wake Field Acceleration (LWFA) large number of
electrons as plasmonic excitation may attract collectively 
protons (or other ions) to
large energies. Furthermore, in the two colliding laser beam wake 
field collider (LWFC) configuration
\cite{LWFC},
the emitted protons are dominantly directed in the direction
of the Electric field of the laser beam, (which is 
linearly polarized). For this we need the LWFC configuration
and the LWFA mechanism for the proton (or ion) acceleration.
The  direction of the Electric field is orthogonal to the
detection of the laser irradiation.

The PIC calculation is performed for electrons, and protons 
accelerated by them. 
The p+11B fusion reaction and the particles created are not 
modeled in this simulation, but according to our experience, 
nanorod antennas accelerate heavier ions similarly. 
The PIC method thus describes the average collective motion of 
the particles. In the 
method we use (and refer to), the method we use, and we usually 
add the local thermal 
MB distribution to this. Since in the present case we want to 
avoid thermalization, 
we approximate the non-thermal distribution with the non-thermal 
CJ distribution 
instead of the thermal MB distribution. Here we only demonstrate 
this, but we do not 
perform the calculation of the two-particle correlation function 
for protons either, 
we only illustrate what the correlation function would be without 
collective flow and 
assuming local thermal equilibrium.

\begin{figure}[!h]
\includegraphics[height=8cm]{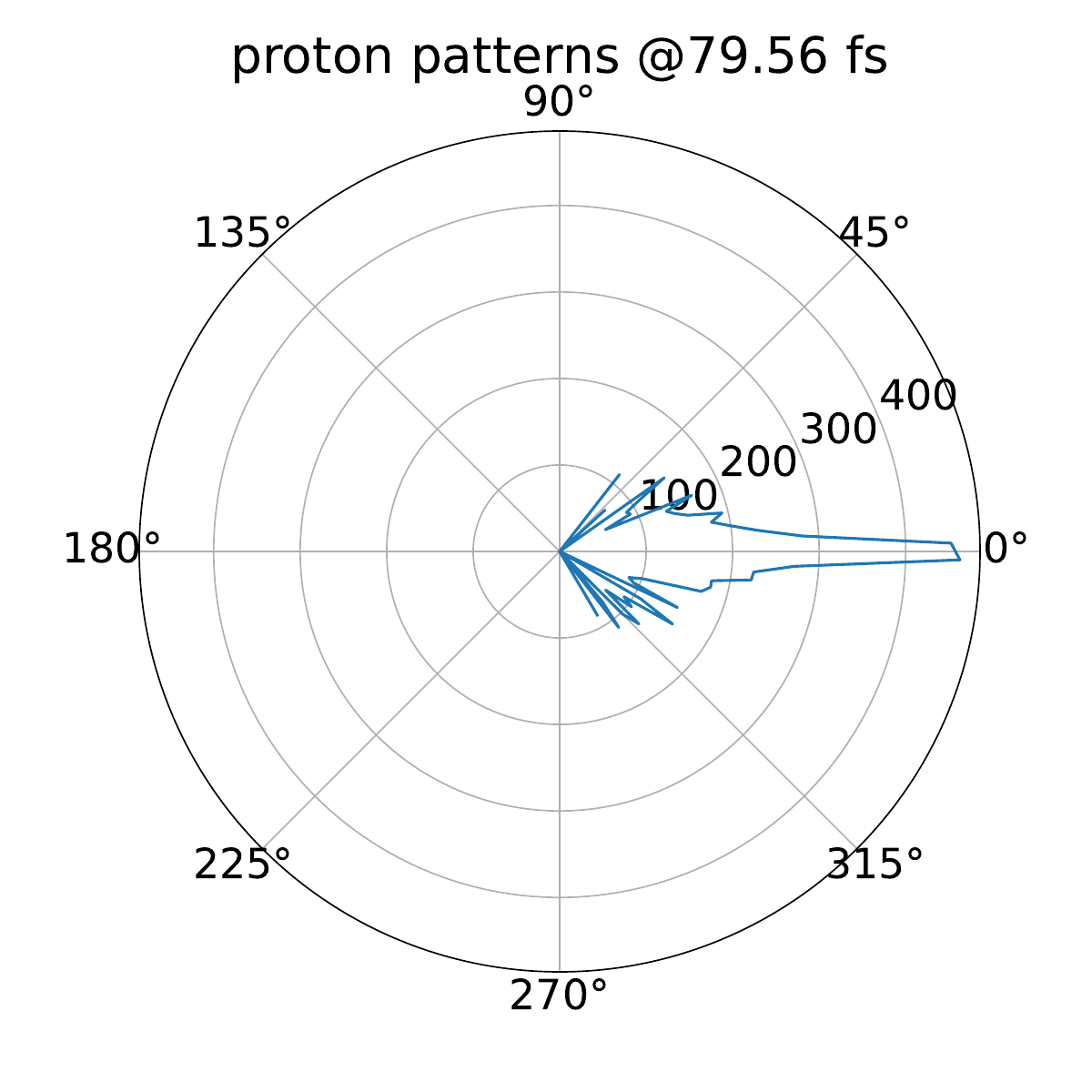}
\vskip -6mm
\caption{
The angular distribution calculated by an EPOCH PIC simulation, of accelerated
protons along a nanorod antenna irradiated by a laser beam.
The time profile of irradaition was a step function 
up to $t=$ 79.56 fs, from {\bf one} side only. 
It was in the $z= 0^o$ direction with a 30mJ laser pulse, with
constant intensity, $I=4\cdot 10^{17}$ W/cm$^2$, 
in the rest frame of the antenna. 
The antenna and the $\vec{E}$ field of the
laser beam point in the orthogonal, $x = 90^o$ direction.
The protons are emitted dominantly in the direction of the laser beam.
The contour line shows the maximum energy of the emitted protons
at the given angles, the scale indicates 100, 200, 300, 400 keV.
The EPOCH code was run for one (85x25 nm) nanorod antenna in a calculational 
box (CB) having a box-size of 530 nm * 530 nm * 795 nm, with 5 nm 
cubic cells, with periodic boundary condition. Electrons in the 
rod were placed randomly with a number density of 5.9e28 electrons/m$^3$. 
The simulation was run for a period of 240 fs.
}
\vskip -4mm
\label{angle}
\end{figure}

The plasmonic nano-rod antennas serve as resonant accelerator modules,
which accelerate neighbouring protons (or ions) with the LWFA
mechanism, Fig. \ref{angle}.
The one-sided irradiation from the left to the right,
i.e. in the $z-$ direction transfers considerable amount
of momentum to the protons and electrons, which is clearly
shown in the $\pm z$ asymmetry of the distribution.

\begin{figure}[h] 
\begin{center}
\includegraphics[width=7.6cm]{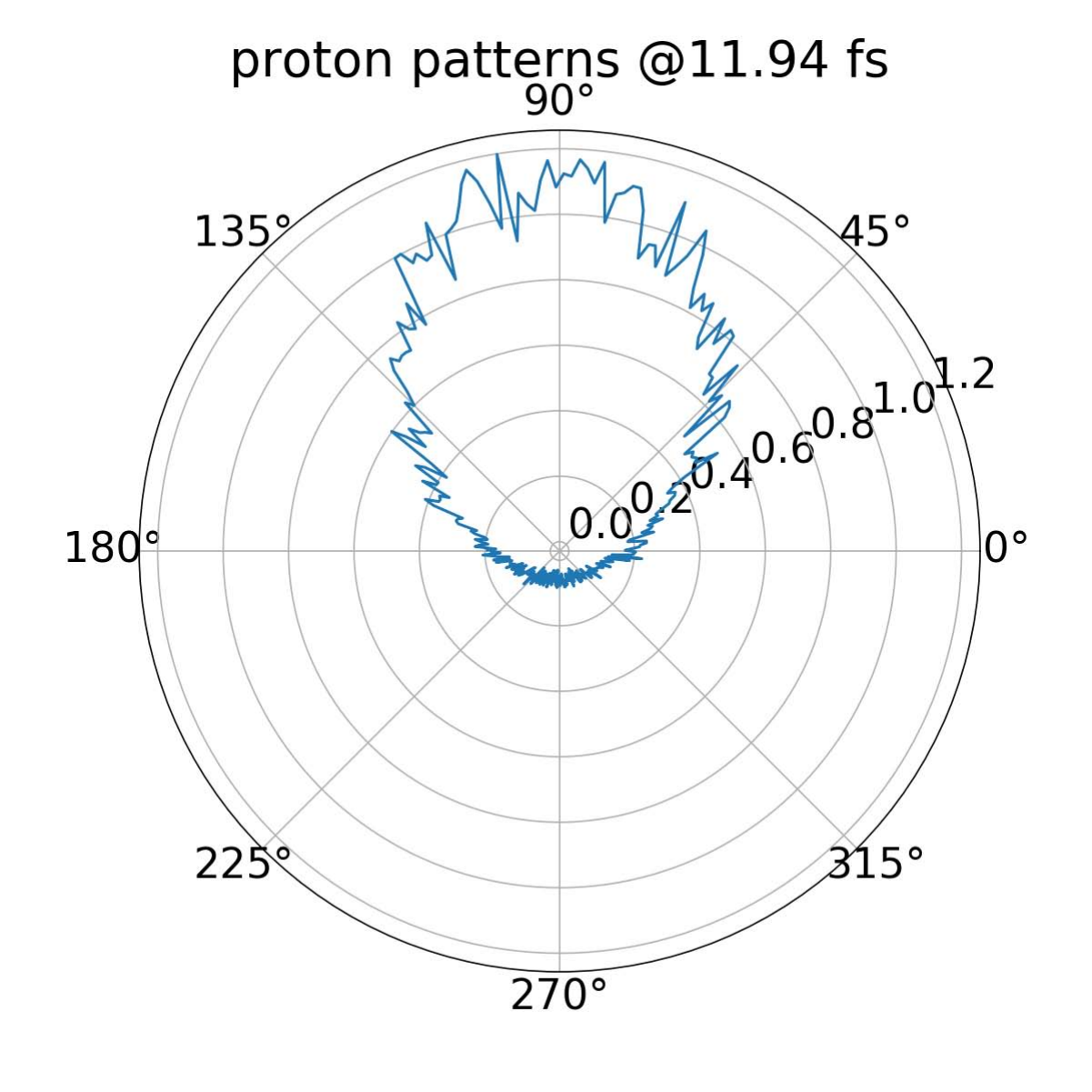}
\end{center}
\vskip -10mm
\caption{The angular distribution, of emitted 
protons along a nanorod antenna
irradiated from {\bf two} sides in the LWFA configuration,
in the $z= 0^o/180^o$ direction with a 30mJ laser pulse with
constant intensity, $I=4\cdot 10^{17}$ W/cm$^2$, with a step
function profile in the rest frame
of the antenna. The distribution is shown one period $T_P$
after the initial transients, $t_o$, i.e. at
$t_o + T_P = 11.94 $ fs after the start if the irradiation.
The antenna and the $\vec{E}$ field of the
laser beam point int the $x = 90^o$ direction.
The outermost contour (1.0 MeV/c) belongs to
momentum of all protons emitted into
a solid angle domain of 4$\pi$/300.
The momentum of the most energetic protons is 13 keV/c.
The number of proton marker particles in the 
EPOCH generated sample is 337058.} 
\vskip -4mm
\label{a18}
\end{figure}

If in the LWFC configuration, irradiate the nanoantenna 
from both sides, $z = 0^o /180^o$, so that the beams
constructively interfere in the middle of the reference frame,
where the nanorod antenna is positioned, the transferred
$z$ -directed momentum is annulled, while the standing waves
in this configuration result in vertical electron and proton acceleration
i.e. in  $\pm x = 90^o /270^o$ directions, Figs. \ref{a18}-\ref{a22}.
The period of the irradiating laser beam is $T_P = 2.65$ fs,
which corresponds to $\lambda = 800$ nm in vacuum. The total irradiation
time for the beam to cross a target of 21 $\mu$m is about 106 fs. 
We assume that the intensity of irradiation is constant in this 
time. 

After the start of the irradiation the electrons start to resonantly
move in the nano-antenna and the protons around the antenna follow
in the LWFC configuration. After an initial transient period of 
$t_o = 9.3$ fs the proton resonance setts in also and this becomes 
noticeable after a couple of periods, $T_P$, Fig. \ref{a18}.
While in the initial few periods the proton energies slowly increase\\
by $t = 18.5$ fs these reach $p=$ 1 MeV/c, \\
by $t = 19.8$ fs these reach $p=$ 1.7 MeV/c, and\\
by $t = 106$ fs these reach $p=$ 12.2 MeV/c.\\
This proton acceleration is achieved by the collective plasmonic excitation
of very large number of correlated electrons.
At the same time the total energy of the accelerated protons
increases much stronger, and by 100 fs it reaches 10 GeV/c,
in a small solid angle range.
%

At the same time the energy of the electromagnetic field in the
Calculation Box surrounding the nanorod antenna decreases by
20-30 \% in 106 fs (see Fig. 3 in ref.
\cite{Papp-Front2023}), 
both in the EPOCH\cite{TDArber2015} and in the COMSOL
model evaluations. This drives the electron plasmonic resonant 
wave as well as the proton resonance and acceleration.

Longer irradiation times increase the amount of protons in the 
same solid angle domain, which becomes increasingly narrower.
At ~ 100 fs the proton momentum in the given solid angle domain
of $4\pi$/300 reaches 10 GeV/c, mainly due to the increased 
number of protons in this domain.  The shape becomes rather 
narrow in the $\pm x$ axis peaking in both directions. 

We present and analyze this method and its results
in a high resolution, Particle In Cell EPOCH kinetic model
\cite{TDArber2015,Epochlink}, 
similarly as it is done for PICR fluid dynamics 
\cite{CsV2014, CsVW2014, hydro2},  
such kinetic models are widely used for laser-plasma interactions 
\cite{Hockney1981,Birdsall2005, Douglass2015, Jiang2014, Wang2015,Papp-23aX}.

In the present simulation we describe the dynamics before and up 
to the moment of ignition. We choose the EPOCH code parameters 
considering the necessary relativistic corrections known from the 
practices in Relativistic Heavy Ion reactions: transitions across 
space-time hypersurfaces and defining marker particles to avoid 
numerical instabilities and considering numerical viscosity.

Fluid dynamics is proven to be the best theoretical method to describe 
collective flow phenomena.
The same model was used to predict the
rotation in peripheral ultra-relativistic reactions
\cite{hydro1},
to point out the possibility of Kelvin Helmholtz Instability (KHI)
\cite{hydro2},
flow vorticity
\cite{CMW12} 
and polarization arising from local rotation, i.e. vorticity
\cite{BCW2013}.
The model was also tested for its numerical viscosity and the
resulting entropy production
\cite{Horvat}.

The two particle correlation studies are used in the field of
ultra-relativistic heavy ion physics for a long time and these
studies were developed to a high level of sophistication determining
not only size and timespan of an emitting source, but its shape, its
dynamics, its expansion and rotation. At CERN the high-resolution 
event by event 4$\pi$ detectors provided a massive amount of data
for two particle correlation studies.  In present,
limited budget laser fusion studies such detector systems are not
always affordable, but one or two detectors are available.

Just as in astrophysics these smaller detector acceptances may provide
sufficient data for important and essential consequences
\cite{Csorgo96,NA44-95}.

Figures Fig. 3, 4, 5 are results of a PIC simulation 
\cite{TDArber2015}
showing the energy and angular 
distributions of the protons accelerated by the resonant nanorods. 
These are non-thermal 
distributions, and with larger number of resonance periods these 
become more and more 
directed towards the axis of the nanorod antenna.

\vskip -4mm

\section{Limits of the EPOCH PIC simulation}

In our EPOCH / PIC simulations we did consider a few types of particle species, 
however, 
these are dependent on the chemical type of the target and that is also changing in our 
experiments. As we are not aiming for DT fusion, these possibilities are numerous and 
could influence the proton distributions. Thus, to estimate the proton angular 
distribution we had to reduce our simulation cases where the proton content of the target 
is dominant, and the other components are assumed to have weaker effects. Unfortunately, 
in the present validation experiments the target is not proton (or Hydrogen) dominated 
but for the dominant UDMA polymer of the 470 nucleons only 38 are Hydrogens. To describe 
the dynamics of all other components is beyond our possibilities and it is also not 
reasonable.  To make this clear for the reader we added a clarifying but somewhat 
shorter explanation. None of the theoretical simulations have so far attempted to 
perform a kinetic simulation for such a complex target chemistry.

At the present type of development in the field of Laser Induced Fusion none of the 
experiments measure any type of two particle correlations. Thus, we do not know what 
possibilities will arise in the future.  From our point of view pre-ignition proton 
distribution is the most important as our goal is to avoid thermalization and 
randomization of protons, in order to lose less energy up to the ignition stage of the 
fusion process.  Obviously, neither we nor any other theoretical groups can simulate 
the whole complexity of the fusion dynamic up to the very end. The experiments are 
doing even less.

The aim of the present article is to estimate the proton acceleration and angular 
distribution up to and at the ignition stage. The detailed dynamics of deuterium and 
alpha production is not aimed at and not addressed by this paper. This is now mentioned 
in the abstract already.  The two-particle method is still applicable for the protons in 
two respects. First the before the full burning of the target fusion fuel and/or in cases
where fusion does not take place, but we have plasma dynamics in the presence of 
resonant nanorod antennas.

\section{Correlation Function}
\label{Corr}

The boson correlation function is defined as the inclusive two-particle
distribution divided by the product of the inclusive one-particle
distributions, such that 
\cite{WF10}:
\begin{equation}
C( p_1 , p_2) =
\frac{P_2( p_1, p_2)}{P_1( p_1)P_1( p_2)},
\end{equation}
where $p_1$ and $p_2$ are the 4-momenta of particles.

To be able to measure such a correlation function we need
a detector (or more), which is able to detect at least two simultaneous 
bosons emitted directly from the fusion ignition, where these were created. 
These particles should not have other interactions or collisions 
(or should not recombine to form atoms or molecules) before detection.

We introduce the center-of-mass momentum
\footnote{The vector $\vec k$ is the wave number vector,
$ \vec k = \vec p/\hbar$ so for numerical calculations we have to
use that $\hbar c =$ 197.327 MeV fm., The same applies to $\vec q$.}
:
$
 k = \frac{1}{2} ( p_1 +  p_2) \ ,
$
and the relative momentum
$
 q =  p_1 -  p_2 \ ,
$
where from the mass-shell condition 
\cite{WF10} 
$q^0 = \vec k \vec q / k^0$.
We use a method for moving sources presented in Ref. 
\cite{Sinyukov-1}.
\begin{equation}
C(k,q) =
1 + \frac{R(k,q)}{\left| \int d^4 x\,  S(x, k) \right|^2}\ ,
\label{C-def}
\end{equation}
\vskip -0.4cm
where`
\vskip -0.6cm
\be
\begin{split}
R(k,q) =&  \int d^4 x_1\, d^4 x_2\, \cos[q(x_1-x_2)] \times\\
& S(x_1,  {k}+{q}/2)\, S(x_2,{k}-{q}/2)\ .
\end{split}
\label{R1}
\ee
Using the emission function $S(x,k)$, 
here $R(k,q)$ can be calculated
\cite{Sinyukov-1, Sinyukov-2, Sinyukov-Cso, Cso-5, CS13}
via the function\\
\be
J(k,q) = \int d^4x\ S(x,k+q/2)\, \exp(iqx) \ ,
\label{J-def}
\ee
and we obtain the $R(k,q)$ function as\\
$ R(k,q) = Re\, [ J(k,q)\, J(k,-q) ] $.

We estimate the local particle (boson) density $n(x)$ 
based on the EPOCH kinetic model using the PIC method.
\footnote{
The net boson density is sufficiently large in macroscopic
experiments, so this approximation is satisfactory. At the present time 
the detection of bosons is relatively limited, so the adequate
measurement technique should still be worked out".  }

For now, we will use the equilibrium single particle 
distribution, $f^J(x,p)$, in
the source functions, i.e. the J\"uttner distribution, which depends 
on the local velocity, $u^\mu(x)$, and we use the notation 
$u_1 = u(x_1) = u^\mu(x_1)$ and temperature $T$.
\be
f^J(x,p) = \frac{n(x)}{C_\pi  (2\pi \hbar)^3  }
\exp\left(\frac{-p^\mu u_\mu (x)}{T(x)}\right),
\label{Jut-2}
\ee
where \ $C_\pi=4\pi m_B^2T K_2(m_B/T)$, \ at temperature $T$,
and $K_2$ is a modified Bessel function and $m_B$ is the boson mass.

{
If at the final stage of the reaction local equilibrium is reached, 
the matter behaves as a fluid and can be characterized by an 
Equation of State (EoS). In this case the emitted particle distribution 
is isotropic and thermal and can be characterized by a the J\"uttner 
distribution or its quantum correspondent. In case of thick target with 
emission from a surface the particle emission is similar to thermal, 
but the emission is in one direction and can be characterized by the 
Cancelling J\"uttner (CJ) distribution. 
}

\begin{figure}[ht] 
\begin{center}
\includegraphics[width=7.6cm]{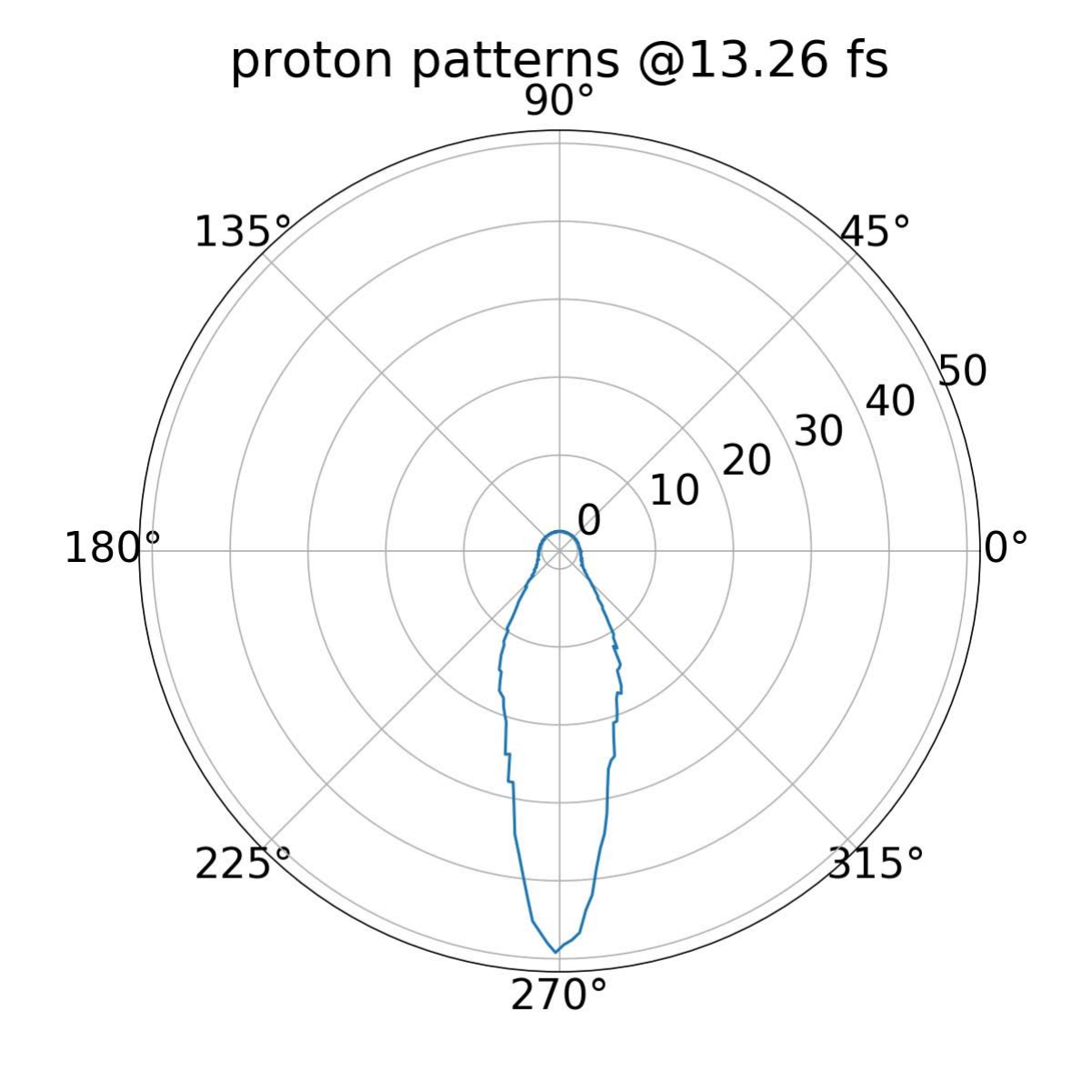}
\end{center}
\vskip -10mm
\caption{The same as Fig. \ref{a18} for
$t_o + 1.5 T_P = 13.26 $ fs after the start if the irradiation,
i.e. half period later. The direction of the motion
of protons is reversed just as the electric field.
The out most contour (50 MeV/c) belongs to 
momentum of all protons emitted into
a solid angle domain of 4$\pi$/300.
The momentum of the most energetic protons is 
36 keV/c, more than earlier in Fig. \ref{a18},
but the increase of the total momentum in the domain
is due to the increased number of the resonating protons.} 
\label{a20}
\end{figure}

The Cancelling J\"uttner distribution, $f_{CJ}$, is defined by 
subtracting the ordinary J\"uttner
distribution (\ref{Jut-2}) with negative velocity, $-v$, from 
original J\"uttner distribution, and
multiplying the obtained result with the step function (Fig. \ref{Tam-F2}):

\begin{figure}[ht] 
\begin{center}
\includegraphics[width=7.6cm]{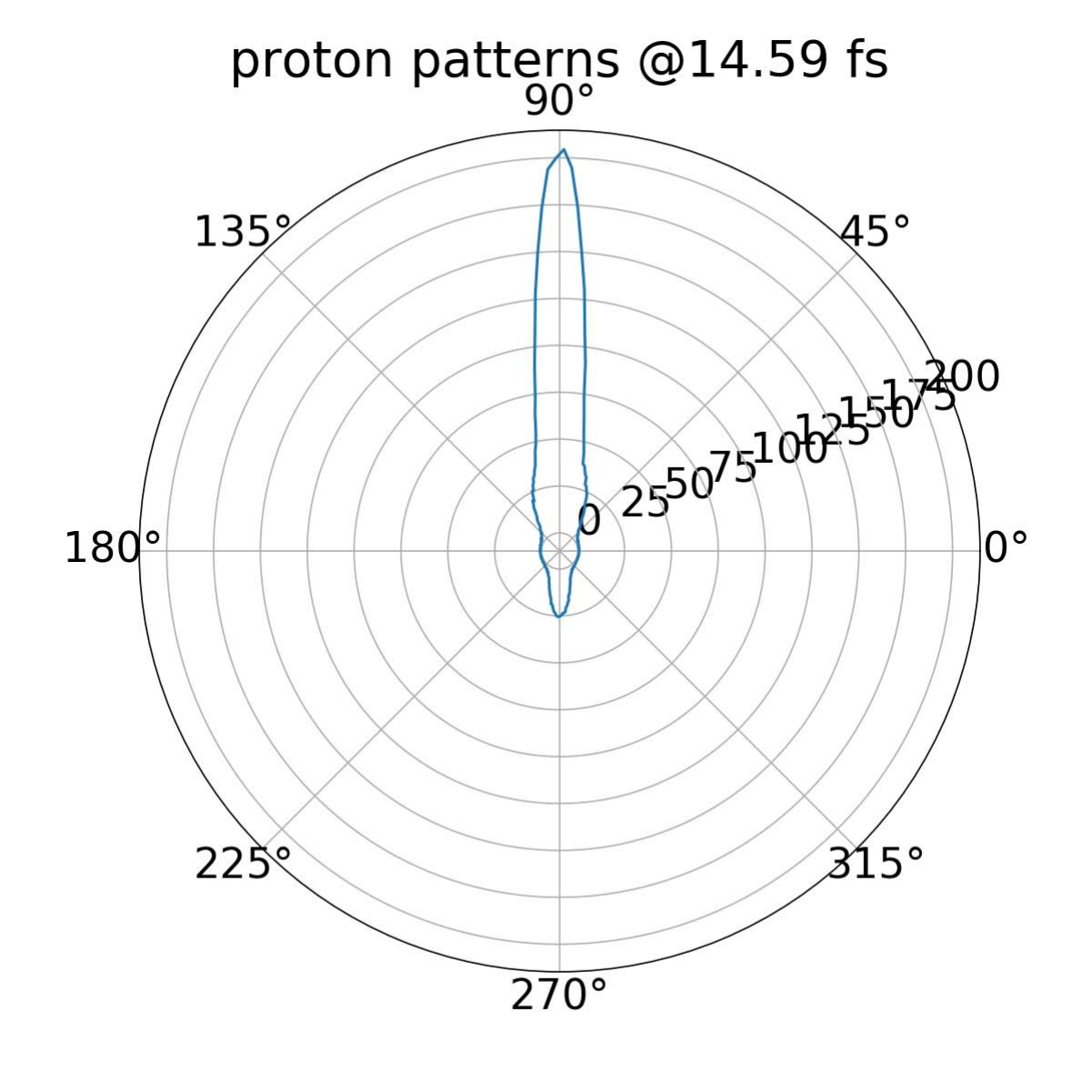}
\end{center}
\vskip -10mm
\caption{The same as Figs. \ref{a18} and  \ref{a20} for
$t_o + 2 T_P = 14.59 $ fs after the start if the irradiation,
i.e. another half period later. The direction of the motion
of protons is reversed again.
The out most contour (200 MeV/c) belongs to 
momentum of all protons emitted into the same
a solid angle domain.
The energy of the most energetic protons is 
68 keV/c, more than before in Fig. \ref{a20},
but the increase of the total momentum in the domain
is due to the increased number of the resonating protons. 
The angular spread of the distribution becomes narrower
at the same time.}
\label{a22}
\end{figure}

Thus, the local invariant particle density is given
by the Cancelling J\"uttner (CJ) distribution
\cite{Tamosiunas}:
\begin{eqnarray}
\lefteqn{f_{CJ}= \frac{}{} 
\frac{\Theta(p^\mu d\sigma_\mu)\,\, n(x)}{C_\pi\, (2\pi \hbar)^3} 
\ \ {\times} }  
\nonumber\\
& & 
\left(\exp{\frac{{-}p^\mu u_\mu ^R}{T}} {-} 
\exp{\frac{{-}p^\mu u_\mu ^L}{T}} \right),
\label{CJ}
\end{eqnarray}
where $d\sigma_\mu$ for a single nanorod is the 
unit normal vector pointing in the 
direction of the nanorod antenna. In case of a full scale macroscopic 
modeling $d\sigma_\mu$ points to the direction of the momentum 
of the marker particle $m_d$ or $m_{He^4}$.
Here 
$ 
u_\mu^R=(\gamma , \gamma v , 0, 0) \ \ 
\textrm{and} \ \ u_\mu^L=(\gamma , -\gamma v , 0, 0)
$ 
in the rest frame of the nanorod or the marker particle.
When $p^\mu d\sigma_\mu = 0$ the function vanishes at the front, even 
without step function $\Theta$.
The role of the $\Theta(p^\mu d\sigma_\mu)$ part is just to eliminate 
the negative part of the distribution.


In recent thick target experiments the photos of measured ejected plume 
\cite{EEE}
showed clear views of asymmetric distributions cut from the left side like 
the Cancelling Jüttner distribution, while the “fitted Maxwell-Boltzmann” 
(Jüttner) distributions did not fit well the data.

The CJ distribution (\ref{CJ}) resembles strongly the 
distribution obtained in the EPOCH  kinetic theory, 
Figs. \ref{a18} (and \ref{a20}).  The increasingly narrower 
distributions can be simulated with increasing velocities $v$.
The "temperature" parameter $T$ in our case represents the
random spread of the kinetic distribution, and this is decreasing
rapidly. In case of Fig. \ref{a18}, $T \sim 13 $ keV, like 
the max proton energy. In case of Fig. \ref{a20}, $T \sim 10 $ keV
approximately one third of the max proton energy.

In case of extreme narrow distributions at late times
as 100 fs, single velocity, non-equilibrated,
{
anisotropic, 
}
distributions can be used instead of the 
Cancelling J\"uttner distributions in the 
two particle correlation studies.  
\medskip

If we assume that the two coincident particles originate from two
points, $x_1$ and $x_2$,
the expression of the correlation function, Eq. (\ref{R1}) will
be become 
\cite{CS13}
\be
\begin{split}
R(k,q) = \int\!  &  d^4 x_1 d^4 x_2\, S(x_1, k) S(x_2, k)
    \cos[q(x_1{-}x_2)] \times  \\
&   \exp\left[ -\frac{q}{2} \cdot \left( \frac{u(x_1)}{T(x_1)}
                                  -\frac{u(x_2)}{T(x_2)}
              \right)\right],
\end{split}
\label{R-def2}
\ee
and the corresponding $J(k,q)$-function is
\be
J(k,q) =  \int d^4x\ S(x,k)\,
\exp\left[ - \frac{q \cdot u(x)}{2 T(x)} \right]\, \exp(iqx)\ ,
\label{J2}
\ee

In Ref. 
\cite{CS13} 
different one, two and four source systems
were tested with and without rotation.
Here we study only the case where the emission is
{\it asymmetric} and dominated by the fluid elements facing the detector.

The modeling for laser induced fusion can be made in two steps.
First, we model one nanorod. In this case the emission is in the
direction of the nanorod, which is orthogonal to the laser irradiation
and points into the direction of polarization of laser light.
Electrons and protons are accelerated in this direction, and
consequently also the deuterium and He$^4$ particles will have
a main direction of emission in this direction. For this stage 
this will be the symmetry axis ($d\sigma^\mu$) pointing in the
$z$ direction. The obtained spread of the distribution can be
fitted to the velocity, $v$ and "temperature", $T$, parameter
of the CJ distribution.

In the second macroscopic step we can identify the marker particles.
These will be our elements of source, $s$, and their direction of
motion will be our symmetry axis, $z$, for the macroscopic step. 

Let us consider the usual conventions, $z$ is the irradiation beam axis, 
and the positive $z$-direction is the direction of the laser beam. 
The laser's polarization vector points into the positive $x$-direction. 
Finally, the $y$-axis is orthogonal to both.
The emission probabilities
from the two source cells of a pair are generally not equal.

The incoming laser beam is not thermal, it is monochromatic 
and linearly polarized, can be transferred in wave guide or 
in coaxial cable. Thematization reduces the average photon 
energy to one third and isotropic and can be converted back to
mechanical or electric energy with significant thermal loss, with
the Carnot efficiency. Thus, we aim to eliminate thermalization 
in the fusion procedure as long as possible. Directed nanorod 
antennas accelerate protons in the direction of the nanorod 
axis, and our goal is to have directed nanorod antenna arrays 
embedded into the fuel target. If in addition, we apply laser 
irradiation from the two opposite sides, but the same (orthogonal) 
polarization, then the kinetic energy of the incoming beams 
is transferred to the nanorod axis direction. Up to this step 
the accelerated proton direction is linear and not thermal, 
the random velocity component is estimated to be minimal. 
(This could be approximated with an anisotropic J\"uttner 
type distribution.)  Validation experiments up to now have 
supported these plans 
\cite{AAA}.

\begin{figure}[!ht]\label{polarJuttner}
\includegraphics[width=75mm, angle=0]{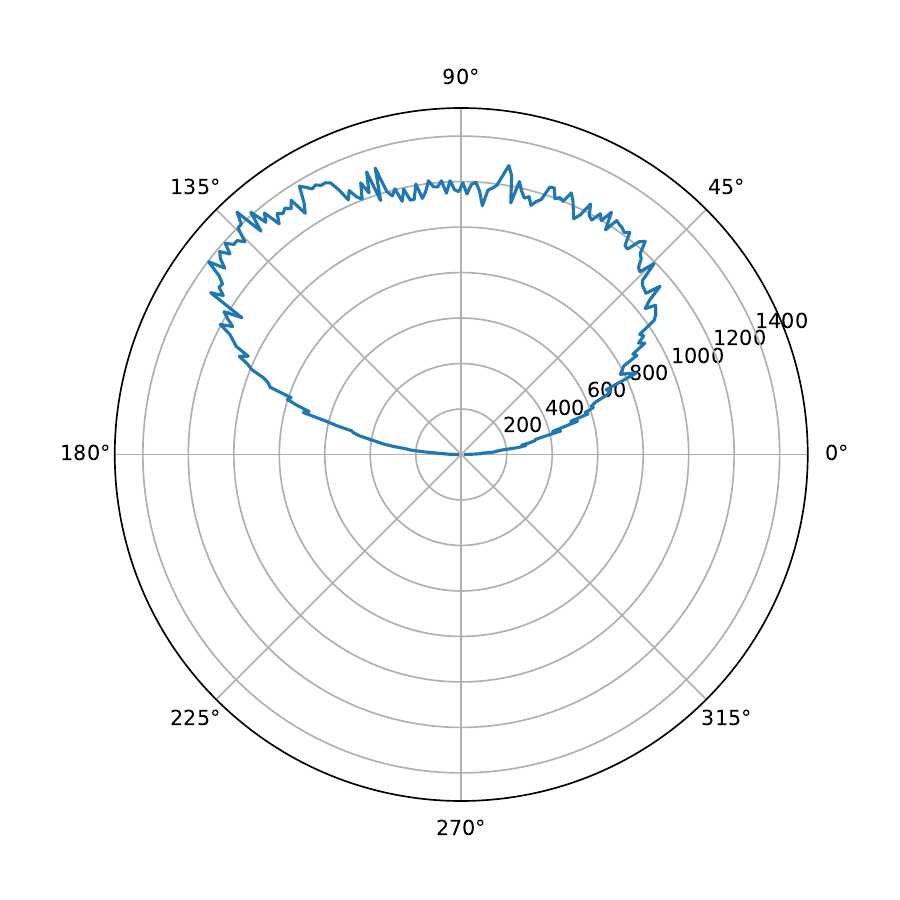}
\caption{
Polar histogram of the Cancelling J\"uttner (CJ) distribution for a 
generated random sample of
300000 protons averaged in 200 bins. The results show a pattern 
resembling features of the EPOCH generated distribution shown in Fig. \ref{a18}. 
However, the parameters of the presented CJ distribution are different. 
The total energy and the temperature parameter in the CJ distribution are 
higher than in the EPOCH simulation. At the same time the ratio of the 
flow velocity to temperature parameters, v/T,  is lower resulting in a 
CJ distribution, which is wider and less elongated than the  
EPOCH simulation.
}
\label{Tam-F2}
\end{figure}

The laser induced fusion of modest pulse energy or intensitu
in a one-dimensional LWFC configuration
needs a target of ~ 20 $\mu$m thick target with a beam diameter
of about 10-30 $\mu$m. 
(Larger pulse energy allows for larger target thickness.)
This is certainly a macroscopic size target
compared to an Event-by-Event relativistic heavy ion collision.
Thus, we investigate the problem from two complementary perspectives: 
first, we analyze the correlation function for a single nanorod using the 
EPOCH kinetic PIC model; second, we explore the system within a dynamical 
framework, treating it as a subsequent independent approach that nevertheless 
aims to capture non-equilibrium and non-thermal features at a 
coarse-grained level.
\vskip -4mm

\section{One Fluid Cell as Source}
\label{SSofcs}

We now assume a source function, 
{
which is reduced to one cell at the moment of emission.
}
Thus, the integration over the 4-volume of an emission
layer is reduced to the 3-volume of a FO hyper surface. For simplicity
of this presentation,
we assume emission along the time like coordinate, $t$, where we assume a
local J\"uttner distribution. Thus, we
have the source function as 
\begin{equation}
S({x}, k) =
G(x)\,H(t) 
\exp\left(-\frac{k_\mu u^\mu (x)}{T(x)} \right)
k^\mu\, \hat\sigma_\mu\ ,
\end{equation}
where $k^\mu \hat\sigma_\mu $ is an invariant scalar, and
$\hat\sigma_\mu $ is a unit vector pointing in the axis, $x$-direction,
of the nanorod antenna. This is the same as the $\vec{E}$ field or
polarization direction of the laser beam.
Furthermore, for the simplicity of presentation we 
assume J\"uttner distribution instead of
Cancelling J\"uttner, and a "temperature" $T$, and velocity $v$. 
Later these can be
replaced by the fitted Cancelling J\"uttner parameters and the 
energy density of the spread of particle emission distribution.

For simple presentation take a single cell and let us 
use a simple quadratic parametrization for $n(x)$ as:
\begin{equation}
G(x) = \gamma n(x) = 
\gamma n_s  \exp\left( - \frac{x^2 + y^2 + z^2}{2 R^2} \right) .
\end{equation}
Here $n_s$ is the average density of the Gaussian source (or fluid cell)
of mean radius $R$.

\bigskip
{\bf Single source at rest:}
The invariant scalar $k^\mu u_\mu$ can be calculated in the frame
where the cell is at rest. 
We have then
\begin{equation}
u^\mu = (1,0,0,0) \Rightarrow 
-\frac{k_\mu u^\mu}{T} =
-\frac{k^0}{T} = 
-\frac{E_k}{T} \ .
\end{equation}
In this simplest case we also assume that the FO direction is
$\hat\sigma^\mu = (1, 0, 0, 0)$, so the $\tau$-coordinate coincides with
the $t$-coordinate, and it is orthogonal to the $x, y, x-$ coordinates.
Then we can make use of the following integral:
\begin{equation}
\int_{-\infty}^{+\infty} e^{-a x^2} d^3 x = 
\left(\frac{\sqrt \pi}{\sqrt a}\right)^3 \ .
\end{equation}
%
%
We can perform the integral along the $t$ direction of
$H(t)$, which gives unity and then
the single particle distribution is
\be
\begin{split}
\int\! d^4x\ S(x, k) = \frac{n_s (k^\mu \hat\sigma_\mu) }{C_n}
\exp\left(-\frac{E_k}{T_s}\right) \times \\ 
\int_{-\infty}^{+\infty} \!\!\!\!\!\!\!\!
                        H(t)  dt 
  \int_{-\infty}^{+\infty} \!\!\!\!
                        e^{-\frac{x^2}{2R^2} } dx 
  \int_{-\infty}^{+\infty} \!\!\!\!
                        e^{-\frac{y^2}{2R^2} } dy 
   \int_{-\infty}^{+\infty} \!\!\!\! 
                        e^{-\frac{z^2}{2R^2} } dz = \\
n_s \ (k^\mu \hat\sigma_\mu) \ \exp\left(-\frac{E_k}{T_s}\right)
    \frac{\left(2 \pi R^2 \right)^{3/2} }{C_n} \ ,
\end{split}
\label{InS}
\ee
here $T_s$ is the {\bf "temperature"} of the source, and $E_k=k^0$
in the rest frame of the fluid cell.
Due to the normalization of $H(t)$ the integral over the time $t$
is unity. The contribution to the nominator from Eq. (\ref{J2}) is
\be
\begin{split}
J(k,q) = \int  d^4x\,  e^{i  q \cdot  x} e^{-q^0/(2T_s)} S({x}, k)  = \\
\frac{n_s\,(k^\mu \hat\sigma_\mu)}{C_n} 
\exp\left[{-}\frac{E_k {+} q^0/2}{T_s}\right] \times\ \ \ \ \\
\int_{-\infty}^{+\infty}\!\!\!\!\!\!\! H(t) e^{iq^0 t} dt
\int_{-\infty}^{+\infty}\!\!\!\!\!\! e^{{-\frac{x^2}{2R^2}}} e^{-iq_x x} dx     \times\ \ \ \    \\ 
\int_{-\infty}^{+\infty}\!\!\!\!\!  e^{{-\frac{y^2}{2R^2}} }
                                         e^{-i q_y y} dy 
\int_{-\infty}^{+\infty}\!\!\!\!\!  e^{{-\frac{z^2}{2R^2}} }  
                                         e^{-i q_z z} dz  = \\
\frac{n_s (k^\mu \hat\sigma_\mu)}{C_n} \left(2 \pi R^2 \right)^{3/2} 
\exp\left[{-}\frac{E_k}{T_s}\right] \exp\left[{-}\frac{q^0}{2T_s}\right] 
\times \ \   \\ 
\exp\left[-\frac{R^2}{2} q^2\right] 
\exp\left[-\frac{\Theta^2}{2} (\hat\sigma^\mu q_\mu)^2\right] \ \ \ \ ,
\end{split}
\ee
where we used
\[
\int_{-\infty}^\infty \exp\left(-p^2 x^2 \pm qx\right) \, dx = \frac{\sqrt{\pi}}{p} \exp\left(\frac{q^2}{4p^2}\right),
\]
as given in \cite[3.323/2]{GR}.
In the time integral the present choice of $\hat\sigma^\mu$ would
give $(q^0)^2$, but we wanted to indicate that other choices are 
also possible and they would yield $(\hat\sigma^\mu q_\mu)^2$.
In the $J(k,q) J(k,-q)$ product the terms $\exp[\pm q^0 /(2T_s)]$
cancel each other.

These integrals can be performed for the Canceling J\"uttner and for
the non-thermal constant velocity distributions also.

Inserting these equations into (\ref{C-def}) we get
\begin{equation}
C( k,  q) = 1 +
\exp\left(-(\Delta \tau)^2 (\hat\sigma^\mu q_\mu)^2 - R^2 q^2\right) \ .
\label{Csst}
\end{equation}
If we have a source point,  which is at longer 
distance from the external side of the source $\Theta$, then
the contribution of the time integral from this point is reduced.

\begin{figure}[ht] 
\begin{center}
\includegraphics[width=7.6cm]{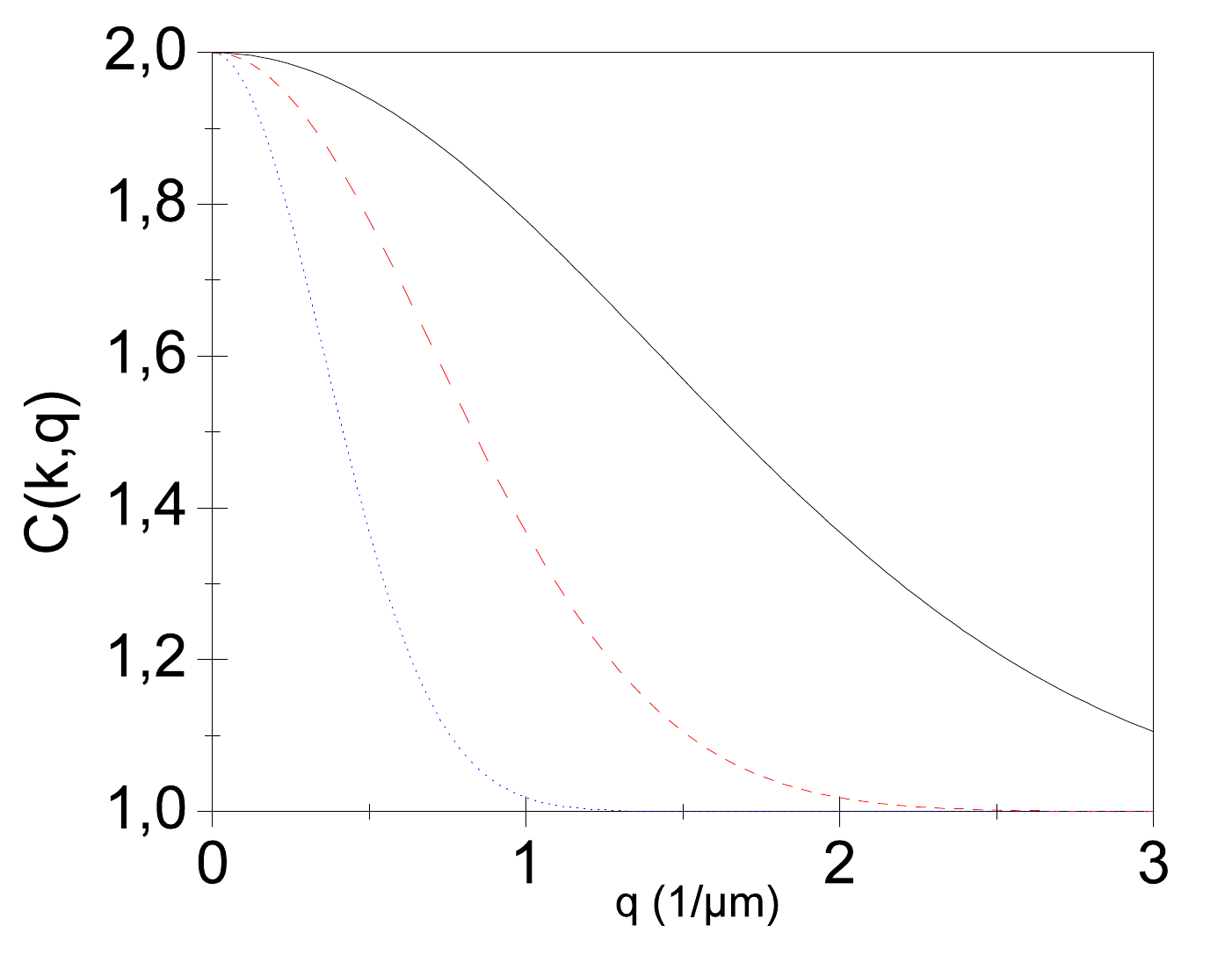}
\end{center}
\vskip -8mm
\caption{ (color online) 
The correlation function, $C(k,q)$, for a single, static, 
spherically symmetric,
Gaussian source with different radii, $R= 4, 1$ and $0.25 \, \mu$m, 
({\it blue dotted, red dashed,} and {\it full black lines} 
respectively), as described by Eq. (\ref{Csss}).
}
\label{F-1}
\end{figure}

If we tend to an infinitely narrow source layer, "sudden full ignition", 
$\Theta \rightarrow 0$, i.e. to a source hyper-surface, then
\begin{equation}
C( k,  q) = 1 +
\exp\left(-R^2 q^2\right) \ .
\label{Csss}
\end{equation}
The $k$ dependence drops out from the correlation function, 
$C$, as the $k$ dependent parts are separable, Fig. \ref{F-1}.

In our EPOCH PIC studies modeling a single nano-rod antenna and its 
surrounding we counted and analyzed a Calculation Box of 
$530 \times 530 \times 795 $ nm.
Thus, a typical radius of such a system is slightly under 1 $\mu$m.

In case of more involved time dependence of the source the
correlation function becomes also more complex, which needs
adequate analysis.

For the study of the rotation of the system the thickness of the
4D ignition layer is of secondary importance, especially if we discuss only
a few fluid source cells. In this case the role of the depth of a source
point within the layer is given by its reduced contribution to the
particle emission. This can be represented much simpler with assigning
emission weights to the small number of sources. Thus, in the following
discussion, we do not go into the details of the time structure of the
emission.

The presented angular emission and femtoscopic studies are not extended 
to full detailed experimental setups. The final plan is to have flat 
target of a thickness and laser pulse duration what the laser light can 
cross once, i.e. for larger pulse energy the target thickness and the 
pulse duration is larger. The laser irradiation will have to be two-sided 
to avoid instabilities and pre-detonation
\cite{DDD}.
During the project development we have initially one-sided irradiation, 
different target materials and target thickness, different nanorod 
antennas and antenna arrays. The angular emission and femtoscopic 
studies will have to repeated for these reaction setups, and the used 
final emitted particle distribution functions should be selected 
according to the actual experimental setup. The three distributions 
mentioned in section II. are the anticipated usual choices.

\section{Outlook}
\label{Outlook}

The two particle correlation measurements for 
laser induced nanoplasmonic inertial confinement fusion 
can be used in two stages. At relatively small laser beam
energy pulses from ~ 30 mJ energy one can see already
nuclear reactions, deuterium production. This is due to the 
increased proton energy caused by the catalyzing effect
of nanorod antennas. Up to now the volume and time
extent of the "Irradiation Volume" was only theoretically 
estimated
\cite{Cs-PRE-23}. 
Recent theoretical analysis indicates that nanorod antennas may
catalyze proton acceleration
\cite{Papp-23aX},
which enables nuclear transmutation and thus fusion reactions.

Two particle correlation analysis with even one or two 
particle detectors can only provide experimental measure
for this intermediate source object dynamically. Up to now
only the crater volume was measured caused by these
nuclear transmutation reactions in the target.

In the future with increased beam intensity and laser
pulse energy at ELI-ALPS we expect energy producing
exothermic $^4$He production, the volume and time extent 
of the formation domain are vital for further development
of the technology based on nanoplasmonics assisted laser
induced fusion.

In case of fusion target with oriented nanorods the angular 
distribution of proton emission can be verified, which confirms
experimentally the directed proton acceleration mechanism
by the applied nanorod catalyzed fusion method.

{
Researchers working in the field of heavy ion collisions can now 
extract information from the correlation function of any two 
particles, not necessarily two identical bosons. Thus, with their 
currently available machinery one can simulate 
e.g. p-p correlation function also. Just as one example 
one can refer to 
\cite{CCC}, 
In contrast to photon correlations in the original HBT method 
we will have to detect other types of emitted particles and the 
final Coulomb interactions among them.

Our general aim is to keep non-thermal processes as long in the 
fusion dynamics as possible. This would allow us to avoid or 
postpone thermalization as long as possible, because thermalization 
leads energy loss. Non-thermal dynamics will allow 
us to construct effective industrial realization of non-thermal 
fusion energy production
\cite{BBB}. 

Still, it is possible that after the nuclear fusion reactions the 
post fusion emitted particles (mainly alphas) will not be fully aligned 
with the nanorod antennas. On the other hand the nanorod antennas
in our directed nano-antenna array target will further accelerate the reaction
products in the same direction, so the non-thermal feature is maintained
or might even be strengthened. 

In our recent measurement at ELI-ALPS 
\cite{AAA}
we have shown that in $p+11B \rightarrow 3 \alpha$ reactions, at 25 mJ laser beam
energy we can accelerate protons to 150 keV energy and achieve fusion. Evaluation 
of the angular distribution of the accelerated protons is in progress. However,
if the protons would have been thermalized their average energy would become about
3 times smaller and would not reach the criotical energy of the 148 keV energy of the
$p+11B \rightarrow 3 \alpha$ resonance.

\bigskip

\noindent
{\bf Acknowledgements:}
Enlightening communication with Zsuzsanna Major is gratefully acknowledged.
L. P. Csernai acknowledges support from
Wigner Research Center for Physics, Budapest (2022-2.1.1-NL-2022-00002).
T.S. Bir\'o, M. Csete, N. Kro\'o, I. Papp, 
acknowledges support by the National Research, Development and
Innovation Office (NKFIH) of Hungary.
We would like to thank the Wigner GPU Laboratory at the
Wigner Research Center for Physics for providing support
in computational resources.
This work is supported in part by
the Frankfurt Institute for Advanced Studies, Germany,
the Hungarian Research Network,
the Research Council of Norway, grant no. 255253,
the Tromsø Research Foundation through Trond Mohn Research 
Foundation’s UiT initiative, FUSENOW (TMF2025UiT01), and
the National Research, Development and Innovation Office of Hungary,
for projects: Nanoplasmonic Laser Fusion Research Laboratory under
project numbers 
(NKFIH-874-2/2020, 468-3/2021, 2022-2.1.1-NL-2022-00002),
Optimized nanoplasmonics (K116362), and
\mbox{Ultrafast} physical processes in atoms, molecules,
nanostructures and biological systems (EFOP-3.6.2-16-2017-00005).


\end{document}